\documentclass[aip,amsmath,amssymb,reprint]{revtex4-1}

\usepackage{graphicx}
\usepackage{dcolumn}
\usepackage{bm}

\usepackage[utf8]{inputenc}
\usepackage[T1]{fontenc}
\usepackage{mathptmx}

\usepackage{hyperref}
\usepackage{color}
\usepackage{soul}

\usepackage{standalone}
\usepackage{pgfplots}
\pgfplotsset{compat=newest}
\usepackage[utf8]{inputenc}
\usepackage{textcomp}
\usepackage{pgfplots}
\usepackage{physics}
\usepackage{amsmath}
\usepackage{braket}
\usepackage{filecontents}

\usepackage{changes}
\definechangesauthor[color=red]{MP}
\definechangesauthor[color=violet]{AB}

\newcommand{\UNIBO}{%
Dipartimento di Fisica e Astronomia, Universit{\`a} di Bologna,
Via Berti-Pichat 6/2, I-40126 Bologna, Italy}

\newcommand{\LPTWON}{%
LP2N, Laboratoire Photonique, Num\'erique et Nanosciences,
Univ. Bordeaux-IOGS-CNRS:UMR 5298, F-33400 Talence, France}

\newcommand{\SOTON}{%
School of Physics \& Astronomy, University of Southampton,
Highfield,  Southampton  SO17  1BJ,  United  Kingdom}

\definecolor{ashgrey}{rgb}{0.7, 0.75, 0.71}

\begin{document}

\preprint{AIP/123-QED}

\title[]{A control hardware based on a field programmable gate array for experiments
  in atomic physics}

\author{A. Bertoldi}
\affiliation{\LPTWON}%

\author{C.-H. Feng}
\affiliation{\LPTWON}%

\author{H. Eneriz Imaz}
\affiliation{\LPTWON}%

\author{M. Carey}
\affiliation{\LPTWON}%
\affiliation{\SOTON}%

\author{D. S. Naik}
\affiliation{\LPTWON}%

\author{J. Junca}
\affiliation{\LPTWON}%

\author{X. Zou}
\affiliation{\LPTWON}%

\author{D. O. Sabulsky}
\affiliation{\LPTWON}%

\author{B. Canuel}
\affiliation{\LPTWON}%

\author{P. Bouyer}
\affiliation{\LPTWON}%

\author{M. Prevedelli}%
\email{marco.prevedelli@unibo.it}
\thanks{The following article has been accepted by Review of Scientific Instruments. After it is published, it will be found at \url{https://publishing.aip.org/resources/librarians/products/journals/}.}
\affiliation{\UNIBO}%

\date{\today}

\begin{abstract}
  Experiments in Atomic, Molecular, and Optical (AMO) physics require
  precise and accurate control of digital, analog, and radio frequency
  (RF) signals. We present a control hardware based on a field
  programmable gate array (FPGA) core which drives various modules via
  a simple interface bus. The system supports an operating frequency
  of 10 MHz and a memory depth of 8 M (2$^{23}$) instructions, both
  easily scalable. Successive experimental sequences can be stacked
  with no dead time and synchronized with external events at any
  instructions. Two or more units can be cascaded and
  synchronized to a common clock, a feature useful to operate large
  experimental setups in a modular way.
\end{abstract}

\maketitle

\section{\label{sec:introduction}Introduction}

Computer control of experiments is a common task of ever increasing
complexity in physics. The implementation of long sequences of events
where different instruments must execute commands at a given time
is mandatory for all but the simplest experiments.

In the past, the number of computer controlled instruments was
relatively small and the type of interfaces was often limited to GPIB
and RS-232. It is now common to require the control of tens of
instruments via USB, Ethernet, PXI, VXI, and others, in addition to
the legacy interfaces mentioned above. The problem of computer control
is aggravated by the need to make frequent changes both to the
experimental sequence and the set of instruments; versatility and
ease of programming are essential features. Moreover, both software
and hardware obsolescence can play a role, since the lifetime of an
experiment is typically longer than the time scale over which operating
systems and computers' architectures evolve. From the software point
of view, the problem has been tackled by developing libraries with a
unified syntax for many interfaces \cite{VISA} and complete frameworks
\cite{TANGO,EPICS}, aimed mostly at large scale experiments in
particle physics.
 
Here, we focus on the specific problem of controlling AMO experiments
where usually outputs are limited to digital and analog signals
changing according to a well defined temporal schedule. By ``well
defined'' we mean that most operations must take place with a strict
temporal resolution, typically of the order of 10 $\mu$s or better for
a cold atoms experimental setup.  It is generally not possible to
fulfil such a constraint with a modern general purpose multitasking
operating system (OS), so dedicated OSs have been developed. Both
commercial \cite{LBRT} and open--source \cite{Rtai_and_Xen} real--time
operating systems (RTOS) are available.  Reasonably complex AMO
experiments \cite{Rosi2014} have been controlled with the RTOS
described in Ref. \onlinecite{Rtai_and_Xen}.  While the software
approach has the advantage of simplicity, versatility and, in case of
open--source solutions, low cost, a hardware one has generally
superior performances in terms of temporal resolution. In the latter
case it can actually be implemented a finite-state machine (FSM)
synchronous with a common master clock (MC), where the digital outputs
of the FSM, interpreted as commands by some auxiliary hardware
modules, are sent out only at state transitions.  The actual execution
time of a command clearly depends on the modules response time, cable
induced delays etc. Often, however, for a MC in the 1 MHz to 10 MHz
range and typical AMO laboratory size experiments, these effects can
be neglected, at least for digital signals, so delay and jitter are
both a small fraction of the MC period. A possible implementation of
the FSM mentioned above is a digital pattern generator (DPG),
i.e. a device that stores in its internal memory a two--column matrix
where the first column represents a time interval expressed in MC
cycles while the second column gives the state that a sufficiently
large number of digital outputs should assume at that transition
time. Often, as in our case, it is more convenient to store not the
absolute time but the interval from the previous transition to
increase the maximum time span of a pattern.

Once the matrix is loaded into the DPG memory, execution can be
started via a software command or an external hardware
trigger. DPGs are commercially available \cite{PCI654x} or have been
implemented in software in dedicated microprocessors running a single
process i.e. a delay loop \cite{Hosak2018}. The abundance on the
market of plug-in modules carrying a FPGA, plus a synchronous dynamic
random access memory (SDRAM) chip, and some kind of fast interface,
usually USB or Ethernet, gives the possibility of implementing high
performance DPGs using a low--cost and easily available, credit--card
size modules. Note that the most recent, high performance, FPGAs might
be available on a module quite before they find their way into full
featured commercial products.

A home--made DPG based on open-source software is more resilient, in
our experience, to the main problem plaguing control hardware setups:
obsolescence. Closed--source drivers for plug-in PC boards might not
be updated for new OS versions after a few years; even the connectors
used to fit the boards in the PC may disappear! The shelf life of a
FPGA--based module is short, but given its limited cost, is very
convenient to stockpile them while available. When, eventually, the
module becomes obsolete, migrating to an updated version requires a
relatively minor effort involving redesigning the printed circuit
board (PCB) hosting the module and porting mostly open-source software
and firmware.

DPGs can be used to drive directly the digital output lines of a
control hardware as suggested in Ref. \onlinecite{Hosak2018}, but
versatility can be added by connecting a DPG to a primitive digital
bus with address and data sections. The DPG acts then as a master
module and performs timed write operations on auxiliary modules
implementing standard functions such as control of digital and analog
outputs (DO and AO respectively) or RF signals (RFO). We define, for
brevity, the combination of DPG and auxiliary modules operating
synchronously with the MC as ``synchronous control hardware'' (SCH).
A PC, possibly running a RTOS, usually takes care of the rest of the
instruments, performing tasks that are not time--critical, using
software timers.

This basic architecture has been adopted by many different groups in
AMO physics and multiple designs have been published in literature
ranging from specific applications \cite{Malek2019} to complete
general purpose designs including both software and hardware
descriptions.

These solutions allow control of complex AMO experiments
developed for diverse research fields ranging from quantum metrology
\cite{Pezz2018}, frequency standards \cite{Ludlow2015}, to quantum
simulation \cite{Georgescu2014}.

Some recent articles published in this journal are, in chronological
order, Refs. \onlinecite{Gaskell2009, Starkey2013, Keshet2013,
  Pruttivarasin2015, artiq2018, Perego2018, Donnellan2019}; together
with references therein, they give a broad review on the
subject. Moreover, various groups report online on their control
systems \cite{SchreckCS_URL, LENS_URL}. As we mentioned above,
obsolescence is often the main limiting factor for these designs so,
usually, any hardware that is maybe ten years old or more is using
some outdated or hard to find components. For example, parts of the
circuit described in Ref. \onlinecite{Gaskell2009} are implemented
with obsolete medium scale integrated circuits, while the control
systems described in
Refs. \onlinecite{Keshet2013,SchreckCS_URL,LENS_URL} employ PC plug-in
and/or PXI cards not in production anymore.

We present here the SCH that we developed for cold atom
experiments in atom interferometry \cite{Canuel2018} and cavity QED
with Bose-condensed gases \cite{Naik2018}, focusing specifically on
our latest DPG, capable of storing up to 8 M$=2^{23}$ instructions
(see \onlinecite{Mega}), controlling up to 128 auxiliary modules and
running with a MC of 10 MHz, which could also be used by itself, as a
replacement for the design discussed in Ref. \onlinecite{Hosak2018}.

The main difference between our design and most of those recently
published is essentially a simpler centralized architecture instead of
a distributed one.

As an example, Refs. \onlinecite{artiq2018,Perego2018,Donnellan2019}
share the same architecture implemented in
Ref. \onlinecite{Gaskell2009}, where every intelligent module (IM)
\cite{IntMod}, i.e. a module implementing a bidirectional fast
communication interface, carrying a FPGA or a CPU, has a comparatively
small first-in-first-out (FIFO) instruction buffer but communicates
with the PC or other modules to receive data even when executing a
sequence of instructions.  In this way the maximum sequence length is
not limited by the FIFO size.
  
Moreover, in order to strictly bound the worst case delay for
arbitrary large systems, each module usually drives only a rather
limited number of outputs, compared to our design. This allows
arbitrary expansion of the control hardware without any increase in
latency.

By comparison we take advantage of the large amount of SDRAM currently
available and its large readout bandwidth to implement a centralized
system with just one IM that will be sufficient to drive enough
outputs to control most AMO experiments. This reduces the complexity
of all the auxiliary modules, since none of them has to be an IM,
without a serious loss of performance.

It should be mentioned that an intermediate solution has been also
adopted\cite{Keshet2013,Starkey2013} by storing the columns of the DPG
matrix in separate devices. A master module stores the first column,
i.e. the temporal data, and sends a pseudoclock signal to the
auxiliary modules. A pseudoclock has a positive edge transition
synchronous to the MC only when any of the auxiliary modules must
change its output. Each auxiliary module, in turn, stores its version
of the second column of the DPG matrix, i.e. the output data, in an
internal FIFO. An excellent discussion of the pros and cons of this
approach can be found in Sec. III of Ref.~\onlinecite{Starkey2013}.

In the following sections we describe in detail the architecture of
the our SCH (Sec. \ref{sec:architecture}) and,
specifically the DPG performance and the RFO modules
(Sec. \ref{sec:characterization}). Finally in Sec.  \ref{sec:further}
we discuss possible improvements.

In order to provide useful material both to those interested in a
complete, out-of-the-box, control system and to those that would like
to modify our work or to integrate parts of it into other available
designs, we provide the complete system documentation
\cite{zenodo2019} including hardware design files and full source code
for the firmware and the software.

\section{\label{sec:architecture}System architecture}

\begin{figure}
\includegraphics[width=.48\textwidth]{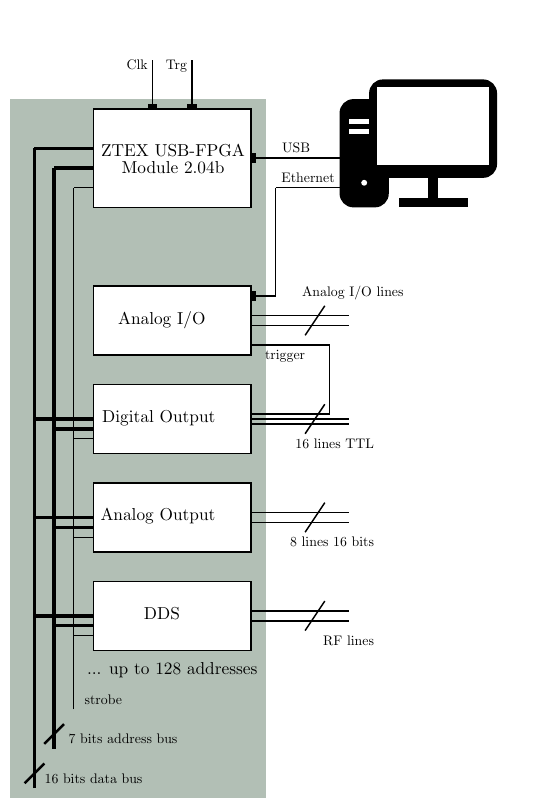}
\caption{\label{fig:scheme} Typical Architecture of the control
  hardware. The master controller is a FPGA card, connected to a PC
  via USB, driving modules via a digital bus. The system can be
  clocked internally or via an external signal connected on the "Clk"
  input; the program execution is controlled either via a software
  trigger from the PC or the "Trg" input. Up to 128 modules providing
  digital, analog, and RF outputs can be connected on the bus. The
  system can provide precise trigger signals to other hardware, here
  represented as a generic analog I/O unit, interfaced to the PC for
  programming and data acquisition.}
\end{figure}

The SCH interfaces with any PC using a USB 2.0 connection and
synthesizes digital, analog, and RF signals for the experimental
setup, as shown in Fig. \ref{fig:scheme}.

The master module stores the instructions received from the PC and,
after receiving a software or a hardware trigger signal, executes
timed write operations on the bus, thus implementing the synchronous
component of the control hardware. The bus has a data width of 16 bits
and an address width of 7 bits. The only control signal is a strobe
line since a master write is the only allowed bus operation.

To our knowledge this bus was originally developed in Paris about 20
years ago\cite{SchreckPhD}, and is still in use at least as reported
in Refs. \onlinecite{SchreckCS_URL} and \onlinecite{LENS_URL}.

Any auxiliary modules latch the content of data bus at the rising edge
of the strobe signal when the content of the address bus matches the
module's address.

Due to the primitive nature of the bus, modules can not return data to
the master. Hence, whenever data must be acquired, some specific
hardware, e.g. a digital oscilloscope or an ADC sampling card, must be
programmed by the PC via an external interface before starting the
program execution and triggered, when required, by the SCH. Finally,
at the end of the experimental sequence, data must be acquired,
processed and stored by the PC before starting the next sequence.

A similar approach, based on triggering dedicated external
instruments, can also be used for generating complex waveforms when it
is impossible or impractical to use a standard AO channel controlled
by the SCH.
In Fig. \ref{fig:scheme}, a module implementing the last two functions
is represented as the generic analog I/O box.

Presently, only one module at a time can be addressed, limiting the
possibility of a simultaneous change of the output lines. A detailed
discussion on this point together with a possible improvement is
deferred to Sec. \ref{sec:further}.
  
\subsection{\label{ssec:DPG}Master module}

The plug-in FPGA module implementing the DPG in the SCH is the
USB-FPGA module 2.04b from ZTEX \cite{ztexURL}. This module includes a
Xilinx Spartan 6 FPGA, interfaced to USB 2.0 via a Cypress USB-FX2
microcontroller, and to a 64 Mbytes, double data rate (DDR) SDRAM chip
via a 16 bits bus capable of running at 200 MHz.  The module hosts a
serial EEPROM to store the FPGA firmware.

The FPGA module is used to control the peripheral cards via a
parallel and write-only bus.
A series of write operations on the bus at specified time intervals
can be downloaded in the board memory and later executed sending
either a software command or a hardware trigger. We call each timed
write operation an \textit{instruction} and a series of instructions a
\textit{program}. The maximum size of a program is 8 M instructions,
limited by the available SDRAM. For reference, a typical program length
in our experiments is of the order of $10^4$ instructions.

The module operates with a MC at 10 MHz, provided either from an
external source or, for simplicity, synthesized from an onboard 24 MHz
quartz of unspecified accuracy. Replacing the quartz with one of known
specifications is a simple operation.  Internally, the FPGA operates
at $f_i=200$ MHz. If $f_c$ denotes the MC frequency a strobe pulse has
a width of $1/2f_c$ and the temporal granularity of the SCH is
1/$f_c$, i.e. 100 ns, which is adequate for most AMO experiments. On
the other hand, the longest programmable interval between two
instructions is $(2^{36}-1)/f_c$, corresponding to about
6872~s. Longer intervals can be programmed by inserting ``dummy''
write operations i.e. instruction not followed by a strobe pulse. When
using the external trigger, the input signal is sampled at $f_i$, to
reduce the risk of missing short pulses, but execution will start at
the next positive edge of the MC, as expected in a SCH.

Since an instruction is coded by 8 bytes in the module's SDRAM, the
required peak bandwidth between the SDRAM and the FPGA is 76.3
Mbytes/s at $f_c=10$ MHz. This is more than a factor 10 smaller than
the peak bandwidth of $800$ Mbytes/s achievable with this module;
increasing $f_c$ up to at least 40 MHz should be therefore possible.

The program is loaded into the module memory through the USB 2.0 bridge
and a parallel interface from the microcontroller and the FPGA, whose
bandwidth is limited at a measured average of about $5 \times 10^5$
instructions/s by the USB interface.

Break points can be inserted in a program after any instruction, in
order to suspend execution with the output lines in a well determined
state for debugging purposes or to wait for an external event. Program
execution can be resumed with a digital signal or with a software
command. This mode of operation can be exploited, for example, to
synchronize the experiment with the power line, to reduce noise, and
to monitor trap loading and trigger at a specified fluorescence
level. A more complex SCH can be implemented by adding a
cascade of ancillary master modules waiting for a trigger sent from an
upper level in order to implement a tree structure similar to that
described in Ref. \onlinecite{Perego2018}. Clearly, in this case, an
external common MC shared by all the modules should be used. This will
be discussed in more detail in Sec. \ref{sec:further}.

\subsection{\label{ssec:peripherals}Auxiliary modules}

Many auxiliary modules have been developed but the most common ones
are of three types: DO, AO, and DDS-based RFO. In addition to these
modules, a general purpose analog I/O module for complex operations is
often needed. Here we provide a quick overview of our most common
modules even if they are not the main focus of this article; the
target is to give an overview of all the hardware necessary for
assembling a control hardware.

All the modules fit on 100 mm$\times$160 mm Eurocard PCBs, to be
mounted in 3U, 19'' racks, and have their addresses programmed with
miniature switches. The master module cannot read from the bus, so the
software must have an \textit{a priori} knowledge of auxiliary
modules' addresses and functions. A more advanced, self-configurating
design, can be implemented, if desired.

The DO modules are just an address decoder and a 16 bits latch so when
writing to a DO module the content of the data bus is stored and
transferred to its 16 output lines. All the lines on the same DO
module can be changed simultaneously while changes on lines on
different modules are separated by at least one MC period.

The AO modules are almost as simple, including a 16 bits DAC per
address, generating an analog voltage, typically in the $\pm 10$ V
range. Parallel input DAC with the inputs connected to the data bus
are particularly convenient and, in order to increase the number of
outputs per PCB, we have recently developed an 8 channel, 16 bits DAC
with parallel input, based on the Texas Instruments DAC8728. This AO
module responds to 8 addresses, one for each channel. Again writing to
different AO channels can not be simultaneous. The minimum delay of 1
MC period is however often negligible when compared with the typical
settling time of a DAC chip. As an example the DAC8728 has a settling
time to 1 LSB for a full range voltage step of about 15 $\mu$s. We
will discuss in Sec. \ref{sec:further} possible schemes to overcome
this limitation.

The RFO modules are based on DDS chips from Analog Devices. Different
chips have been adopted over time, depending on specific
applications. Our most commonly used chip has been the AD9958, which
has two independent DDSs channels clocked at 400 MHz and can generate
RF signals up to 160 MHz with a frequency resolution of about 0.1
Hz. The output includes a 10 bit linear amplitude control. The maximum
output power is +10 dBm.

For higher frequency and better resolution we have recently developed
a module based on the AD9912 chip, a single channel DDS is capable of
a maximum clock frequency of 1 GHz, producing a RF output up to 400
MHz with a resolution in the few $\mu\mathrm{Hz}$ range. Amplitude
control is provided by a programmable logarithmic attenuator with a
dynamic range in excess of 30 dB in 0.25 dB steps (Analog Devices
HMC759). The module output power is again at least +10 dBm.

We will refer in the following to the single channel and the dual
channel modules as RFO1 and RFO2 respectively.

The RFO modules are the most complex modules and pose some challenges
for inclusion in the SCH. For a precise arbitrary frequency ramp, some
chips, i.e. the AD9954 or the AD9910, include an internal volatile
memory where a table of arbitrary frequencies can be stored and later
recalled automatically with a programmable delay between consecutive
steps. The special case of a linear frequency or amplitude ramp is
commonly implemented in hardware in many DDS chips, like the AD9958,
so it is often sufficient to program the start and stop values, the
step size and the dwell time at each point.

In both cases, however, a rather complex programming sequence is
needed. Even a discrete frequency and amplitude step might require
sending more than ten bytes to a DDS chip. Most chips have only a
serial interface, so some logic is required to convert the 16 bits
read from the data bus, decode them, and finally send the proper
programming sequence to the DDS. This introduces a delay, between the
moment when the command is issued and the one when the RF appears at
the output, which can be large when compared with $1/f_c$. Moreover,
16 bits do not contain enough information to fully specify frequency
and amplitude even for a single RF channel. There are two possible,
simple solutions.

The first consists of assigning a separate address for frequency and
amplitude for each RF channel and mapping, linearly, the values on the
data bus to the frequency range actually required for a specific
application. As an example, consider driving an acousto-optic
modulator operating at its central frequency, in this case of $f_0=80$
MHz. The -3 dB bandwidth of the modulator is typically limited to
$f_0\pm 30$ MHz so the $2^{16}$ values on the data bus could specify,
a frequency in the interval 50--110 MHz with a frequency resolution of
the order of 1 kHz, which is adequate for most applications.

The second possibility is to use the content of the data bus as a
pointer to a lookup table (LUT) stored in memory on the RFO module,
where the complete state of a DDS driving RF channel is stored. The
total number of combinations of frequency and amplitudes
typically used in a given experimental sequence is a small fraction of
all the possible values given the DDS resolution. This offers versatility
at the cost of the increased complexity of generating and sharing the LUT
tables between the modules and the PC.

In both cases, the module must include memory and processing
capabilities. The two natural choices are a FPGA or a
microcontroller. A FPGA will allow driving the DDS serial interface at
a speed close to the maximum bandwidth, minimizing the output delay
but with greater complexity and cost. A microcontroller, on the other
hand, will provide a simple and cheap solution with a larger delay.

Our RFO modules include a microcontroller with USB interface
(Microchip PIC 18F2550) and use the LUT model.  The LUT is loaded and
updated in the microcontroller's flash memory from the PC via
USB. Both RFO modules share the same PCB hosting the microcontroller
and the bus interface while the DDS and the other RF components are on
a separate daughter--board. This modular structure simplifies the
upgrades necessary when better DDS chips become available.

The typical delay between a write operation and the RF output change
is of the order of a few tens of $\mu$s, specifically 30 $\mu$s per
channel when changing both frequency and amplitude with the RFO2
module and about 25 $\mu$s for the same operations for the RFO1
module.

When this sort of delay is not acceptable we can adopt a partial
workaround: most DDS chips have double-buffered registers, meaning
that the new frequency, amplitude etc. are preloaded in buffers and
transfered to the DDS core only after pulsing a dedicated line. We
usually have the microcontroller applying the pulse as soon as the
data are sent, but we have the option to delay the pulse with a second
write operation on the bus. This means that we can change the RF
output synchronously with bus write operations as soon as we allow
enough time, as specified above, between two consecutive state
changes. Note that for the RFO1 module this method does not apply
since the amplitude control is implemented with an external
attenuator.

More advanced, albeit more complex, FPGA--based DDS designs have been
published in Refs. \onlinecite{Pruttivarasin2015,Perego2018}.

Except for the "Trg" and "Clk" lines (see Fig. \ref{fig:scheme}), this
control hardware does not provide integrated input lines, either
digital or analog, given the unidirectional nature of the system
bus. The possibility to acquire signals is implemented with external
hardware synchronized by DO lines: for example, the open--source
STEMlab 125-14 module \cite{redPitayaURL} can provide 2 inputs sampled
at up to 125 MHz, with 50 MHz analogue bandwidth and 14 bit amplitude
resolution, and auxiliary analogue inputs at 250 kHz with 12 bit
resolution. The STEMlab board provides analog output lines with
comparable bandwidth and resolution, and 16 general-purpose
input/output (GPIO) lines. A STEMlab module can be triggered at the
desired time by a DO line, while communication with the PC
controlling the experiment uses a 1 Gbit/s Ethernet interface, as
shown in Fig. \ref{fig:scheme}.  Typically this module provides the
analog input (AI) function for i.e. sampling the fluorescence signal
of a detection photodiode. The board can process, store and finally
notify the PC that the data acquired are available. For convenience we
have prepared a simple PCB hosting the STEMlab board to derive its
supply and to be mounted in a 19'' rack with the other modules.

\subsection{\label{ssec:software}Software}

There are a few different software layers in a control system. In
our experience, trying to use open-source or, at least, free software
at every layer minimizes obsolescence and portability
problems. Hardware and software are as independent as possible in this
system, to allow for partial or incremental upgrades.

The firmware required by the auxiliary modules, notably the
aforementioned RFO modules, includes the code for the microcontroller
written in C and compiled with an open-source compiler for small
devices, SDCC \cite{SDCC} and some simple Verilog code for a complex
programmable logic device, (CPLD) that provides the interface to the
bus. The LUT size, stored in the microcontroller flash memory, is 16
kbytes, sufficient for storing, for example, 2048 different
combinations of amplitude and frequency for the RFO1 module or 1024
combinations of frequency, phase and amplitude for each channel of the
RFO2 module.

We have actually developed two different firmware versions for the
RFO2 module. The first one, simpler and sufficient for most
applications, only sets phase, frequency and amplitude for each
channel. The second takes full advantage of the chip's capabilities,
allowing linear frequency (amplitude, phase) ramps and frequency-shift
(amplitude-shift, phase-shift) keying modulation up to 16 different
values.

For the RFO2 module, therefore, specifying values for the LUT is
complex enough that we have chosen to implement a description based on
the extensible markup language (XML) format, while for the RFO1 a
simple text file with frequency-amplitude pairs is used.

The master module requires firmware for the FPGA and for the
microcontroller controlling the USB interface.

The master module actually comes with an open-source default firmware
for the microcontroller, implementing a high speed bridge from the USB
to the FPGA that we found adequate for our purposes. The code can be
easily modified and recompiled with SDCC if required.

The firmware for the FPGA is written in Verilog and compiled using the
free version of the design software provided by Xilinx \cite{ISE}.
The code uses, whenever convenient, proprietary black-boxes that are
freely available but not open-source. Specifically in this design we
take advantage of the SDRAM controller and the FIFO generator. For a
migration to full open-source code, we remark that the most complex
module is the SDRAM controller and it could be replaced with designs
from open-source repositories such as Ref. \onlinecite{opencores}.

The Verilog code includes a main FSM (MFSM) that accepts and
executes the commands sent to the FPGA and, finally, replies to the PC
when appropriate, a memory interface and the MC-synchronous FSM (SFSM).

The commands follow a simple custom protocol, documented in detail in
Ref. \onlinecite{zenodo2019}. Here we provide just a quick
overview. Every command is formed by two ASCII characters followed by
parameters when required. Commands fall in the following four
categories: trigger control, execution control, memory access and
status request.  Trigger control commands select between an internal
or an external trigger source.  Execution control commands arm the
trigger in external mode or start/resume execution in internal
mode. There are also two stop commands for ending execution after the
next instruction or forcing an immediate stop, respectively. This last
command is useful since, as mentioned above, the next instruction
could be scheduled after more than 6800 s.

Instructions can be loaded one at time in memory in random order. This
is less efficient than grouping many consecutive instructions in a
single packet but, on the other hand, it allows arbitrary incremental
changes on a program already resident in memory, simplifying multiple
executions of programs where only some instructions are changed
i.e. for parameter scanning. Due to this choice the command
interpreter can not easily determine if a program is complete so a
specific command must be sent after loading the last instruction.

Finally a status request command returns to the PC a packet of data
specifying the MFSM status (idle, waiting for trigger,
running, stopped at a break point). In the last two cases also the
current instruction number is returned.

Presently polling the master module via the status request command is
the only way for the PC to determine when a program has terminated due
to the USB bus nature (communication can be initiated only by the USB
master i.e. the PC). A possible improvement on this behaviour will be
discussed in Sec. \ref{sec:further}.

As soon as a complete program is loaded, the memory interface starts
reading from the SDRAM in bursts of four 16 bits-wide words, so that
each burst corresponds to a single instruction, and feeds data to the
the input of a dual port FIFO 64 bits-wide with a depth of 512
locations inside the FPGA. The FIFO acts as buffer to provide data to
the SFSM even when the SDRAM is busy performing refresh
cycles. The combination of FIFO depth and average SDRAM bandwidth
guarantees, as shown in Sec. \ref{sec:characterization}, that the
maximum execution rate of an instruction per MC cycle can be sustained
indefinitely.

The SFSM fetches the next instruction from the FIFO just
one MC cycle before the current instruction is executed and performs a
sequence of operations that, as a first step, involve splitting the
instruction in its four components\cite{zenodo2019}: the 7 bits
address, the 16 bits data, the 36 bits time interval and the 5 control
bits. Presently only 3 control bits are used. They are used to mark
the last instruction, set a break point and, finally, enable the
generation of the strobe pulse. Note that not firing the strobe pulse
can be useful both inserting dummy instructions in order to increase
the maximum time interval between instructions and to disable the
execution of few selected instructions for testing.  

The main control program runs on the PC and it is the last software
layer. It is responsible for taking a "user friendly" description of
the program, usually a timeline of each output, and translating it
into a sequence of instructions for one or more DPGs.  The concept of
"user friendly" is so highly subjective, that has led to a myriad of
different solutions. Here, we mention the most recent ones of which we
are aware, namely
Refs. \onlinecite{Keshet2013,Starkey2013,Perego2018}. We point out
that while Ref. \onlinecite{Keshet2013} uses a graphics library that
runs only under the Windows OS, Ref. \onlinecite{Starkey2013} is
written in Python and Ref. \onlinecite{Perego2018} is written in C++,
and both use a graphics library available for the most common
OSs. 

\begin{figure}
\includegraphics[width=.48\textwidth]{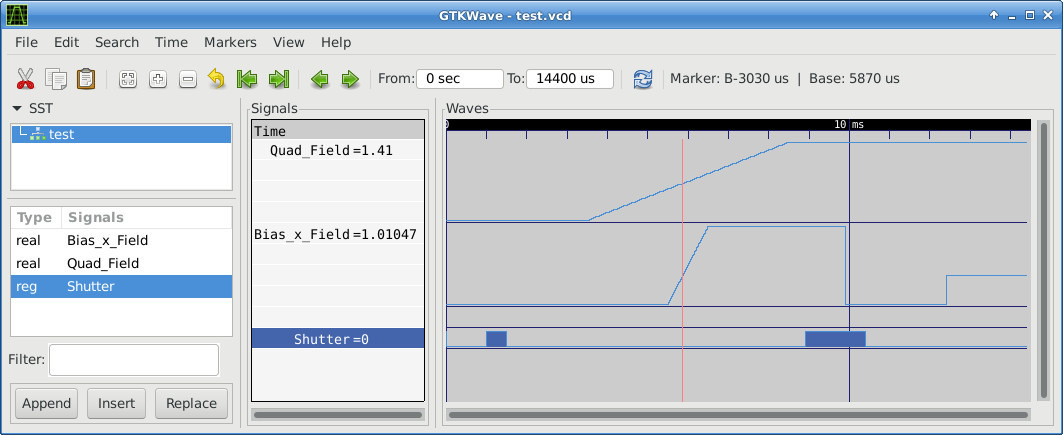}
\caption{\label{fig:gtkwave} Screenshot of a VCD file viewer
  \cite{gtkwave}. The traces shown are similar to Fig.~2 from Ref.
  \onlinecite{Starkey2013}, for comparison, and show the timeline of
  the current in two electromagnets and the status of a
  shutter. The viewer offers many possibilities to customize the
  output format and cursors for measuring times, time differences,
  and the amplitudes of analog signals.
}
\end{figure}

We are confident that adding support for our SCH to both the programs
mentioned above should be relatively simple however have ported to our
DPG two different programs that were previously developed for
different hardware, one written in C and the other in Python,
following, respectively, the two approaches described above. The
source code for both of them is available\cite{zenodo2019}.

A useful feature that we included in both codes is the possibility to
write a file with a graphical representation of the program regardless
of the way in which it was originally generated, in order to have a
quick visual check of the experiment: the IEEE standard 1364-2001
\cite{vcd} specifies a format for the output of hardware simulations
called value change dump (VCD) for which a full-featured open-source
viewer is available \cite{gtkwave}. A VCD file representing a program
is a very helpful tool to check what a program
does. Fig.~\ref{fig:gtkwave} shows, as an example, the screenshot of
the VCD viewer applied to an experimental sequence involving three
signals: the current in two electromagnets and the state of an optical
shutter, controlled by a digital output line. The sequence is similar
to that reported in Fig.~2 in Ref. \onlinecite{Starkey2013}.

Adopting a control program for an experiment is a choice that
ultimately depends on the end--user and will inevitably change over
time. We think that a wise choice relies on open-source code that can
run on multiple OSs. When designing a master module, the best solution
is to use a simple set of commands that can be implemented using as
many programming languages as possible, letting the users choose the
control program that best suits their needs. As we have seen, our
master module communicates through USB packets implementing the
commands briefly discussed above, in order to be usable with a broad
range of commercial software programs, such as Labview, or programming
languages.

\section{\label{sec:characterization}Characterization / Performances}

\begin{figure}[t]
\includegraphics[width=.5\textwidth]{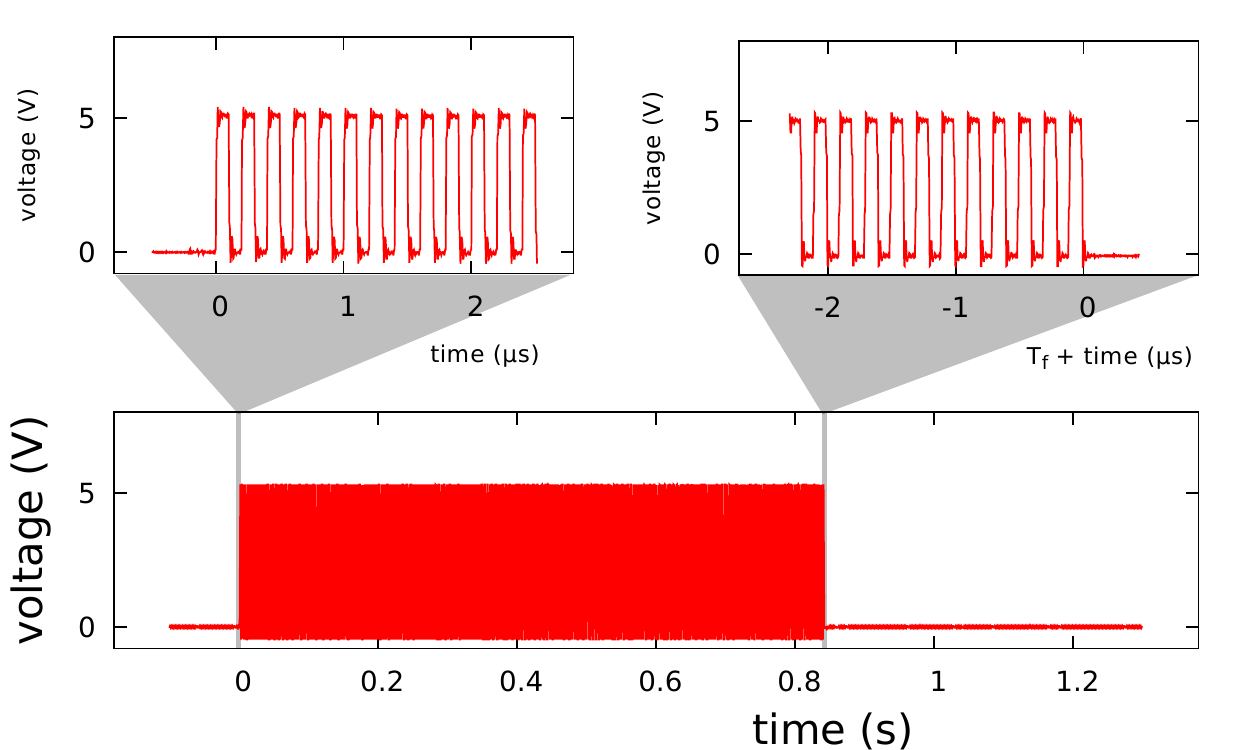}
\caption{A digital output line is toggled every 100 ns, i.e. at the
  full rate given by the 10 MHz signal used to clock the system, for 8
  M $= 2^{23}$ times (see \onlinecite{Mega},). The 5 MHz square wave
  burst periodicity of the commuting digital output is maintained for
  the whole sequence, as highlighted by the insets showing the
  beginning and the end of the program. The time for the final section
  of the sequence is referred to the falling edge of the last pulse.}
\label{fig:8M}
\end{figure}

The modular design of the control hardware allows for easy
customization of the configuration to meet the specific requirements
of the experiment, in terms of the number and type of output
signals. Typical instruments that are controlled directly or through
specific drivers by such systems include: optical shutters, acousto-
and electro-optic modulators, electromagnets, electronic switches, and
CCD cameras.  Here we would like to focus mainly on the
characteristics of our SCH.

Two features of our master module are noteworthy, namely its ability
to operate at full speed i.e. to send a command at every MC transition
and the possibility to pause and re--trigger using external signals in
the same program, at will.

The transmission bandwidth between the SDRAM and the FPGA is
sufficient to maintain the peak speed of one instruction per MC cycle,
indefinitely at $f_c=10$ MHz. This is shown in Fig.~\ref{fig:8M},
where a burst of 8 M transitions of a digital output of a DO module,
limited only by the SDRAM size, is shown.

Synchronization of the program execution is critical in AMO
experiments.  The measurement precision can be improved by
synchronizing part of the experimental sequence with external signals,
as discussed in Sec.~\ref{ssec:DPG}. Moreover, complex setups composed
of independently controlled, spatially separated sub-units require a
precise synchronization to achieve optimal performances.

\begin{figure}[t]
  \includegraphics[width=.5\textwidth]{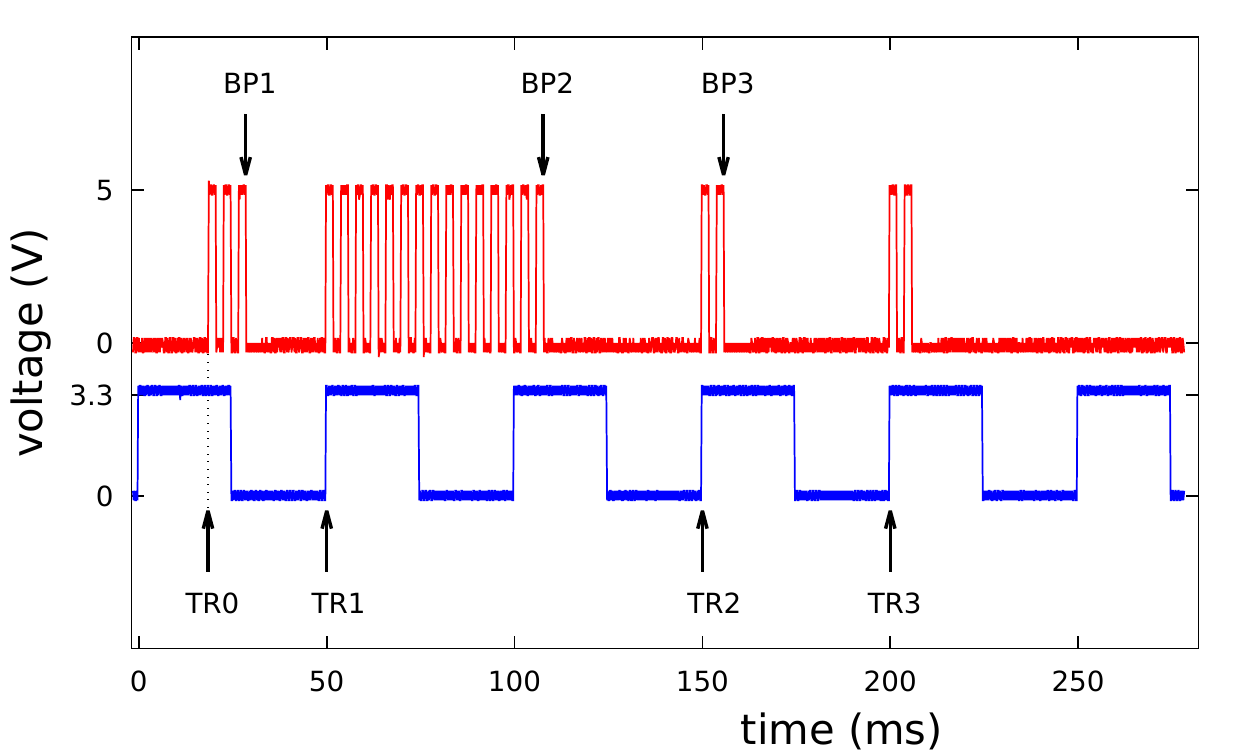}
  \caption{\label{fig:bp} The sequence in the red trace is triggered
    via software at a time indicated by TR0 and shows the digital output
    commuting every 2 ms, hitting 3 break points, after commuting 6,
    30, and 4 times, respectively. The three break points are
    indicated by the labels BP$n$. The sequence resumes on
    the positive edges of the blue trace, a square wave at 20
    Hz, at the times labelled by TR$n$. After TR3, the sequence ends with 4 last commutations of the digital line.}
\end{figure}

Since break points can be inserted in the program at any instruction
and the execution resumed via a software command or hardware trigger
using a dedicated "Trg" digital input (see Fig. \ref{fig:scheme}), it
is simple to synchronize DPGs and relative modules to a master DPG
unit. In Fig. \ref{fig:bp}, we show an example of break points with
hardware retriggering. The toggling of a digital output line (red
signal) is suspended three times, and each time execution resumes at
the first positive edge of the Trg input (blue line), in this case
provided by a 20 Hz square wave. Since the output will always be
synchronous with the MC and an extra MC cycle is internally used to
change state, the worst case latency between the rising edge of Trg
and the resuming of execution will be $2/f_c$ or 200 ns in this
example. This is, of course, how a SCH is supposed to work and a
reduction in latency can only be achieved by increasing $f_c$.

We have discussed above why, due to the synchronous nature of our
hardware, presenting delay and jitter measurements of the DO and AO
modules will provide limited information. The delay is dominated by
cable length for the DO modules and by DAC settling time for AO
modules while jitter is expected to be negligible when compared with
$1/f_c$.
  
It is instead worth showing the use of double-buffered registers in
the AD9958 DDS to implement limited strictly synchronous capabilities
of the RFO2 module.

Normally a command to the RFO2 module uses the first 10 bits on the
data bus to specify an entry in the 1024 elements LUT table. The
frequency and amplitude for each channel corresponding to that entry
-- the phase is not usually changed -- are loaded into the the DDS
chip by the microcontroller via a serial interface. As mentioned
above, this takes about 30 $\mu$s per channel. As soon as the DDS
registers are loaded, the microcontroller pulses a dedicated line
actually forcing the new frequency and amplitude to appear at the RF
outputs with a delay of few tens of ns. Both the presence of the
update pulse and the action of programming the DDS are independently
enabled by separate bits on the data bus. New values can be preloaded
in the DDS without forcing an update and, later, an update pulse can
be applied, without reprogramming the DDS, to change the output in
synchronously to a given MC transition.

\begin{figure}[t]
  \includegraphics[width=.5\textwidth]{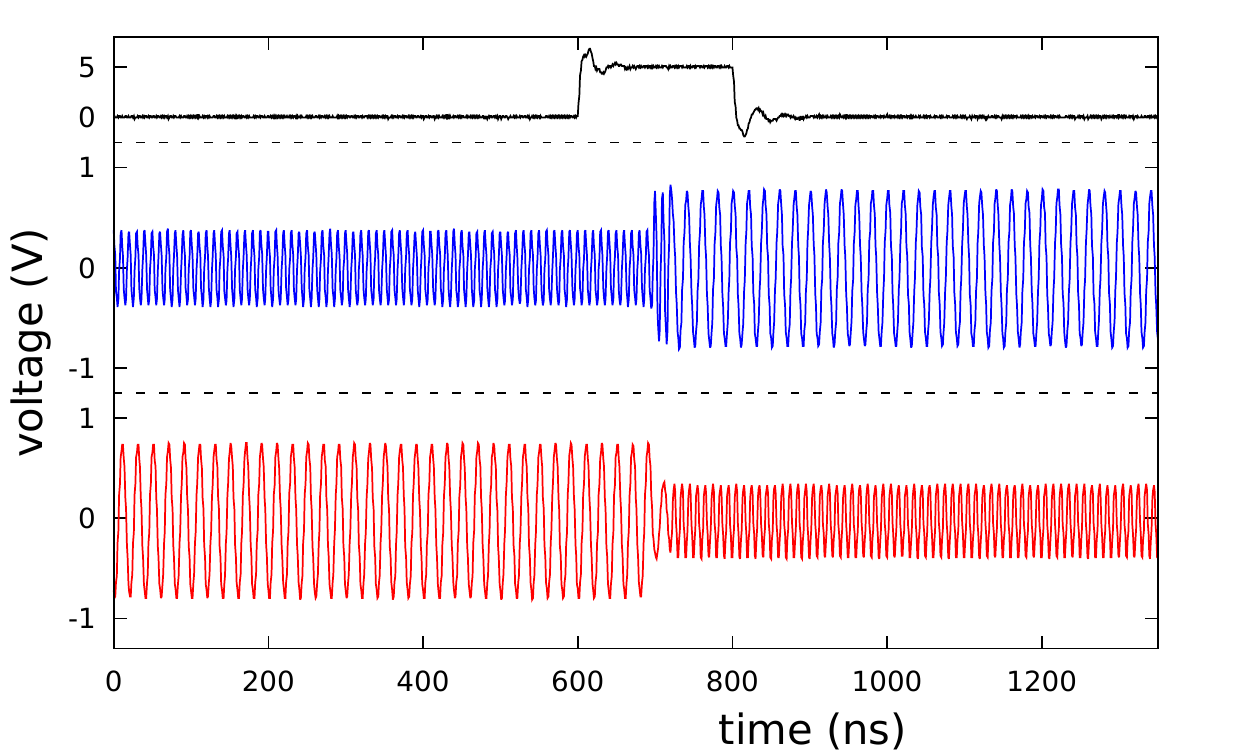}
  \caption{\label{fig:dds} The RF outputs of a dual channel DDS module
    initially at 100 MHz, half power (blue trace) and 50 MHz, full
    power (red trace), exchange frequency and amplitude upon receiving
    an update command from the master module. For reference a digital
    line (black trace) generates a 200 ns pulse, centered around the
    update command. The amplitude change occurring few tens of ns
    before the phase-continuous frequency change is consistent with
    the DDS specifications \cite{AD9958_ds}.}
\end{figure}

This is shown in Fig. \ref{fig:dds} where the two channels of a RFO2
module initially at 100 MHz, half power (blue) and 50 MHz, full power
(red), exchange frequency and amplitude upon receiving an update
command from the master module. The pulse is not shown but, for reference,
the black trace shows an output line of a DO module generating a pulse
with the rising and falling edges one MC cycle before and after,
respectively, the update command.

The RF ouputs change amplitude and frequency well within 100 ns,
meeting the requirements for synchronous operation. Note that the
amplitude changes before the frequency. This is in agreement with the
DDS datasheet \cite{AD9958_ds} where the minimum latencies for
amplitude and frequency are specified as 17 and 29 internal clock
cycles. Since the DDS is clocked at 400 MHz, the expected latencies
are about 42.5 ns and 72.5 ns respectively so amplitudes are expected
to change about 30 ns before frequencies. These latencies are limited
by the DDS itself therefore compare favorably even with those reported
using high speed FPGA to DDS interfaces \cite{Pruttivarasin2015}.

The update rate is however limited to about 16 kHz by the time
required to pre--programm the DDS registers. Faster rates can be
obtained using frequency-shift (amplitude-shift) keying modulation if
only one between frequency and amplitude must be changed and no more
than 16 different values for a single channel, are required.  

The outputs of the RFO2 module share the same clock so keep a definite
phase relation when generating the same frequency. The phase offset
$\Delta \phi$, however, depends from the previous history of the
channels. It is possible to set $\Delta \phi$ to zero by resetting
simultaneously the phase accumulators of both channels inside the DDS
chip. Later $\Delta \phi$ can be set to any desired value in the
$[0,2\pi]$ interval with a resolution of 14 bits.  The option to set
$\Delta \phi=0$ is not implemented in the current firmware version but
can be add, if required.

\section{\label{sec:further}Further developments}
The DPG performances can be readily improved by adopting a more
powerful FPGA module. This also shows how our approach can tackle
obsolescence.  The ZTEX 2.13a, from the same
manufacturer,\cite{ztexURL} is essentially pin-compatible but uses
more recent and powerful components i.e. a Xilinx Artix 7 FPGA and a
256 MBytes DDR3 SDRAM at 400 MHz, bringing the number of instructions
to 64 M and doubling SDRAM peak bandwidth. As a test we have ported,
compiled and simulated the Verilog firmware for this
module\cite{zenodo2019} but, so far, we have not acquired a physical
unit for testing. Note that the ZTEX 2.13a requires a different set of
development tools\cite{FX3,Vivado}. Using this module pushing $f_c$ to
at least 50 MHz should be relatively simple.

Although most experiments will probably need only one DPG module, as
we mentioned above, multiple modules can be cascaded and synchronized
using break points either for convenience or necessity.

As an example of the first, consider auxiliary modules distributed on
19'' racks in different parts of a laboratory. Carrying around long
flat cables for distributing the 24 lines of the SCH bus rather than
just 2 BNC cables carrying the MC and a DO line from the master DPG
rack acting as a trigger would be both unpractical and possibly prone
to noise, especially at high $f_c$.

In the case of the underground, 150 m long atom-laser antenna MIGA
\cite{Canuel2018}, a demonstrator we are developing for atom
interferometry based gravitational wave detection, multiple DPGs will
be instead mandatory. There different atom interferometers, with their
electronic subsystems, will be separated by a distance of the order of
100 m so the issue of achieving synchronous operation when the
transmission delay is large with respect to the MC period must be
addressed.

We have not included in the DPG firmware a syntax for implementing
loops which are instead ``unrolled'' by the PC code before sending a
program to the master module. Due to the large amount of memory
available and the high communication bandwidth, presently we do not
consider this inefficiency as a limiting factor. A more evoluted
firmware can nevertheless be developed.

By using a distribution amplifier and cables with matched lengths,
sufficiently coherent copies of the MC can be distributed among the
various subsystem, then the propagation delay between the master DPG
and each auxiliary DPG must be determined with an uncertainty better
than one MC cycle. Using either BNC cables or optical fibers this
corresponds to an uncertainty, in length, that, at $f_c = 10$ MHz,
should be small when compared to 20 m. The master DPG then must send a
trigger pulse with the correct timing to obtain a distributed SCH.  

If simultaneous change of outputs across different DO and AO modules
is required, multiple solutions are possible. First, as described
above, multiple DPG modules could be used.

As an alternative, with a single DPG, it is possible to reserve a
specific address for sending trigger commands. Any DO or AO module
decoding both their address and the trigger address could interpret
then the content of the data bus as 16 independent trigger lines. This
approach requires, however, the design of a new set DO and AO boards
where the bus interface is implemented using CPLDs rather than
standard TTL chips.

An obvious way to achieve rapid switching of analog outputs is using
auxiliary modules containing some $2^n$-to-one analog multiplexers,
where $2^n$ analog outputs and $n$ digital outputs are required for a
single, fast switching, analog output among $2^n$ preset values. The
cost in term of hardware resources is compensated by a switching time
limited not by the DAC but by the multiplexer.

As we mentioned before, a DPG module with only a USB interface has the
disadvantage that it cannot initiate communication. Moreover including
an Ethernet interface to our SCH will help integration with other
systems. Both issues can be solved by adding a simple auxiliary
credit--card size computer with USB and Ethernet interfaces and
digital input signals capable of generating interrupts, acting as
bridge and, possibly, as protocol translator between the DPG and other
hardware. The DPG could use one of the DO lines to fire an interrupt
in the auxiliary computer, which will forward a service request to the
main PC via Ethernet. An obvious application could be notify program
termination instead of relying on polling.
  
Even without hardware changes there is room for software improvements
that, mostly, could improve reliability. A simple example is given by
the handling of break points. Presently the software in the PC can
easily determine, by polling, how long the DPG has been waiting at a
break point. It could be useful to set a sensible timeout policy to
resume execution or abort the program and set the experiment in a safe
condition.

\section{\label{sec:conclusions}Conclusions}

We have presented a control hardware developed for experiments in AMO
physics; it is broadly applicable to experimental science. The system
provides up to hundreds of digital, analog, and RF signals controlled
with a time resolution of 100 ns, with a wealth of possible
configurations. The program execution can be synchronized with
external events at multiple times in the same sequence. Several
improvements can be easily adopted to the system: the time resolution
could be increased by a factor of 5 to 20 ns by changing the main
clock frequency and with simple modifications of the firmware; the
maximum number of instructions in a program could be increased by a
factor of 4 by adopting a different FPGA board. The modular nature of
the system makes it easy to adapt parts of it, especially the low-cost
DPG module, to other applications and make it more resilient against
inevitable obsolescence. In general, the system could be engineered to
be compatible with portable or space applications, in terms of weight,
volume, and power consumption.

\begin{acknowledgments}

  This work has been partly founded by the ``Agence Nationale pour la
  Recherche" (grant EOSBECMR \# ANR-18-CE91-0003-01, grant ALCALINF \#
  ANR-16-CE30-0002-01, grant MIGA \# ANR-11-EQPX-0028), Laser and
  Photonics in Aquitaine (grant OE-TWC), Horizon 2020 QuantERA ERA-NET
  (grant TAIOL \# ANR-18-QUAN-00L5-02), and the Aquitaine Region
  (grants IASIG-3D and USOFF). M.C. acknowledges support by DSTL-DGA.
  M.P. would like to thank for useful discussions and constant support
  M. De Pas, M. Giuntini and A. Montori at the electronic workshop
  of the European Laboratory for Nonlinear Spectroscopy (LENS),
  Firenze.
\end{acknowledgments}

\appendix

\nocite{*}
\bibliography{main2}

\begin{thebibliography}{37}%
\makeatletter
\providecommand \@ifxundefined [1]{%
 \@ifx{#1\undefined}
}%
\providecommand \@ifnum [1]{%
 \ifnum #1\expandafter \@firstoftwo
 \else \expandafter \@secondoftwo
 \fi
}%
\providecommand \@ifx [1]{%
 \ifx #1\expandafter \@firstoftwo
 \else \expandafter \@secondoftwo
 \fi
}%
\providecommand \natexlab [1]{#1}%
\providecommand \enquote  [1]{``#1''}%
\providecommand \bibnamefont  [1]{#1}%
\providecommand \bibfnamefont [1]{#1}%
\providecommand \citenamefont [1]{#1}%
\providecommand \href@noop [0]{\@secondoftwo}%
\providecommand \href [0]{\begingroup \@sanitize@url \@href}%
\providecommand \@href[1]{\@@startlink{#1}\@@href}%
\providecommand \@@href[1]{\endgroup#1\@@endlink}%
\providecommand \@sanitize@url [0]{\catcode `\\12\catcode `\$12\catcode
  `\&12\catcode `\#12\catcode `\^12\catcode `\_12\catcode `\%12\relax}%
\providecommand \@@startlink[1]{}%
\providecommand \@@endlink[0]{}%
\providecommand \url  [0]{\begingroup\@sanitize@url \@url }%
\providecommand \@url [1]{\endgroup\@href {#1}{\urlprefix }}%
\providecommand \urlprefix  [0]{URL }%
\providecommand \Eprint [0]{\href }%
\providecommand \doibase [0]{http://dx.doi.org/}%
\providecommand \selectlanguage [0]{\@gobble}%
\providecommand \bibinfo  [0]{\@secondoftwo}%
\providecommand \bibfield  [0]{\@secondoftwo}%
\providecommand \translation [1]{[#1]}%
\providecommand \BibitemOpen [0]{}%
\providecommand \bibitemStop [0]{}%
\providecommand \bibitemNoStop [0]{.\EOS\space}%
\providecommand \EOS [0]{\spacefactor3000\relax}%
\providecommand \BibitemShut  [1]{\csname bibitem#1\endcsname}%
\let\auto@bib@innerbib\@empty
\bibitem [{VIS()}]{VISA}%
  \BibitemOpen
  \href@noop {} {}\bibinfo {note} {{Virtual Instruments Software
  Architecture}~(VISA):
  \href{http://www.ivifoundation.org/specifications}{www.ivifoundation.org/specifications}.
  Both commercial (\href{https://www.ni.com/visa}{www.ni.com/visa}) and public
  domain (\href{https://github.com/hgrecco/pyvisa}{github.com/hgrecco/pyvisa},
  limited to GPIB, USB and RS-232 so far) implementations are available. For
  the Linux OS an effort to a unified approach to data acquisition is
  represented by Comedi (\href{https://comedi.org}{comedi.org}). {W}e mention
  commercial products at the sole purpose of identification for helping the
  reader to reproduce our setup. We do not endorse any specific product.
  Products with similar or even better performances might be
  available.}\BibitemShut {Stop}%
\bibitem [{TAN()}]{TANGO}%
  \BibitemOpen
  \href@noop {} {}\bibinfo {note} {{Tango:}
  \href{https://www.tango-controls.org}{www.tango-controls.org}.}\BibitemShut
  {Stop}%
\bibitem [{EPI()}]{EPICS}%
  \BibitemOpen
  \href@noop {} {}\bibinfo {note} {{Experimental Physics and Industrial Control
  System}~(EPICS):
  \href{https://epics-controls.org/}{epics-controls.org}.}\BibitemShut {Stop}%
\bibitem [{LBR()}]{LBRT}%
  \BibitemOpen
  \href@noop {} {}\bibinfo {note} {As an example, National Instruments offers a
  real--time module for the Labview programming environment:
  \href{https://www.ni.com/en-us/shop/select/labview-real-time-module}{www.ni.com/en-us/shop/select/labview-real-time-module}
  .}\BibitemShut {Stop}%
\bibitem [{Rta()}]{Rtai_and_Xen}%
  \BibitemOpen
  \href@noop {} {}\bibinfo {note} {Two real--time open source extensions for
  Linux are RTAI (\href{https://www.rtai.org}{www.rtai.org}) and Xenomai
  (\href{https://xenomai.org}{xenomai.org}).}\BibitemShut {Stop}%
\bibitem [{\citenamefont {Rosi}\ \emph {et~al.}(2014)\citenamefont {Rosi},
  \citenamefont {Sorrentino}, \citenamefont {Cacciapuoti}, \citenamefont
  {Prevedelli},\ and\ \citenamefont {Tino}}]{Rosi2014}%
  \BibitemOpen
  \bibfield  {author} {\bibinfo {author} {\bibfnamefont {G.}~\bibnamefont
  {Rosi}}, \bibinfo {author} {\bibfnamefont {F.}~\bibnamefont {Sorrentino}},
  \bibinfo {author} {\bibfnamefont {L.}~\bibnamefont {Cacciapuoti}}, \bibinfo
  {author} {\bibfnamefont {M.}~\bibnamefont {Prevedelli}}, \ and\ \bibinfo
  {author} {\bibfnamefont {G.~M.}\ \bibnamefont {Tino}},\ }\bibfield  {title}
  {\enquote {\bibinfo {title} {Precision measurement of the {N}ewtonian
  gravitational constant using cold atoms},}\ }\href {\doibase
  10.1038/nature13433} {\bibfield  {journal} {\bibinfo  {journal} {Nature}\
  }\textbf {\bibinfo {volume} {510}},\ \bibinfo {pages} {518} (\bibinfo {year}
  {2014})}\BibitemShut {NoStop}%
\bibitem [{PCI()}]{PCI654x}%
  \BibitemOpen
  \href@noop {} {}\bibinfo {note} {See, e.g., the National Instruments series
  of PCI boards
  (\href{https://www.ni.com/en-us/shop/select/digital-waveform-device}{www.ni.com/en-us/shop/select/digital-waveform-device})
  or the PulseBlaster series at Spin Core
  (\href{https://spincore.com/products/\#pulsegeneration}{spincore.com/products/\#pulsegeneration}).}\BibitemShut
  {Stop}%
\bibitem [{\citenamefont {Ho{\v{s}\'a}k}\ and\ \citenamefont
  {Je{\v{z}}ek}(2018)}]{Hosak2018}%
  \BibitemOpen
  \bibfield  {author} {\bibinfo {author} {\bibfnamefont {R.}~\bibnamefont
  {Ho{\v{s}\'a}k}}\ and\ \bibinfo {author} {\bibfnamefont {M.}~\bibnamefont
  {Je{\v{z}}ek}},\ }\bibfield  {title} {\enquote {\bibinfo {title} {Arbitrary
  digital pulse sequence generator with delay-loop timing},}\ }\href {\doibase
  10.1063/1.5019685} {\bibfield  {journal} {\bibinfo  {journal} {Rev. Sci.
  Instrum.}\ }\textbf {\bibinfo {volume} {89}},\ \bibinfo {pages} {055103}
  (\bibinfo {year} {2018})}\BibitemShut {NoStop}%
\bibitem [{\citenamefont {Malek}\ \emph {et~al.}(2019)\citenamefont {Malek},
  \citenamefont {Pagel}, \citenamefont {Wu},\ and\ \citenamefont
  {M\"{u}ller}}]{Malek2019}%
  \BibitemOpen
  \bibfield  {author} {\bibinfo {author} {\bibfnamefont {B.~S.}\ \bibnamefont
  {Malek}}, \bibinfo {author} {\bibfnamefont {Z.}~\bibnamefont {Pagel}},
  \bibinfo {author} {\bibfnamefont {X.}~\bibnamefont {Wu}}, \ and\ \bibinfo
  {author} {\bibfnamefont {H.}~\bibnamefont {M\"{u}ller}},\ }\bibfield  {title}
  {\enquote {\bibinfo {title} {Embedded control system for mobile atom
  interferometers},}\ }\href {\doibase 10.1063/1.5083981} {\bibfield  {journal}
  {\bibinfo  {journal} {Rev. Sci. Instrum.}\ }\textbf {\bibinfo {volume}
  {90}},\ \bibinfo {pages} {073103} (\bibinfo {year} {2019})}\BibitemShut
  {NoStop}%
\bibitem [{\citenamefont {Pezz{\`{e}}}\ \emph {et~al.}(2018)\citenamefont
  {Pezz{\`{e}}}, \citenamefont {Smerzi}, \citenamefont {Oberthaler},
  \citenamefont {Schmied},\ and\ \citenamefont {Treutlein}}]{Pezz2018}%
  \BibitemOpen
  \bibfield  {author} {\bibinfo {author} {\bibfnamefont {L.}~\bibnamefont
  {Pezz{\`{e}}}}, \bibinfo {author} {\bibfnamefont {A.}~\bibnamefont {Smerzi}},
  \bibinfo {author} {\bibfnamefont {M.~K.}\ \bibnamefont {Oberthaler}},
  \bibinfo {author} {\bibfnamefont {R.}~\bibnamefont {Schmied}}, \ and\
  \bibinfo {author} {\bibfnamefont {P.}~\bibnamefont {Treutlein}},\ }\bibfield
  {title} {\enquote {\bibinfo {title} {Quantum metrology with nonclassical
  states of atomic ensembles},}\ }\href {\doibase 10.1103/revmodphys.90.035005}
  {\bibfield  {journal} {\bibinfo  {journal} {Rev. Mod. Phys.}\ }\textbf
  {\bibinfo {volume} {90}},\ \bibinfo {pages} {035005} (\bibinfo {year}
  {2018})}\BibitemShut {NoStop}%
\bibitem [{\citenamefont {Ludlow}\ \emph {et~al.}(2015)\citenamefont {Ludlow},
  \citenamefont {Boyd}, \citenamefont {Ye}, \citenamefont {Peik},\ and\
  \citenamefont {Schmidt}}]{Ludlow2015}%
  \BibitemOpen
  \bibfield  {author} {\bibinfo {author} {\bibfnamefont {A.~D.}\ \bibnamefont
  {Ludlow}}, \bibinfo {author} {\bibfnamefont {M.~M.}\ \bibnamefont {Boyd}},
  \bibinfo {author} {\bibfnamefont {J.}~\bibnamefont {Ye}}, \bibinfo {author}
  {\bibfnamefont {E.}~\bibnamefont {Peik}}, \ and\ \bibinfo {author}
  {\bibfnamefont {P.}~\bibnamefont {Schmidt}},\ }\bibfield  {title} {\enquote
  {\bibinfo {title} {Optical atomic clocks},}\ }\href {\doibase
  10.1103/revmodphys.87.637} {\bibfield  {journal} {\bibinfo  {journal} {Rev.
  Mod. Phys.}\ }\textbf {\bibinfo {volume} {87}},\ \bibinfo {pages} {637--701}
  (\bibinfo {year} {2015})}\BibitemShut {NoStop}%
\bibitem [{\citenamefont {Georgescu}, \citenamefont {Ashhab},\ and\
  \citenamefont {Nori}(2014)}]{Georgescu2014}%
  \BibitemOpen
  \bibfield  {author} {\bibinfo {author} {\bibfnamefont {I.}~\bibnamefont
  {Georgescu}}, \bibinfo {author} {\bibfnamefont {S.}~\bibnamefont {Ashhab}}, \
  and\ \bibinfo {author} {\bibfnamefont {F.}~\bibnamefont {Nori}},\ }\bibfield
  {title} {\enquote {\bibinfo {title} {Quantum simulation},}\ }\href {\doibase
  10.1103/revmodphys.86.153} {\bibfield  {journal} {\bibinfo  {journal} {Rev.
  Mod. Phys.}\ }\textbf {\bibinfo {volume} {86}},\ \bibinfo {pages} {153--185}
  (\bibinfo {year} {2014})}\BibitemShut {NoStop}%
\bibitem [{\citenamefont {Gaskell}\ \emph {et~al.}(2009)\citenamefont
  {Gaskell}, \citenamefont {Thorn}, \citenamefont {Alba},\ and\ \citenamefont
  {Steck}}]{Gaskell2009}%
  \BibitemOpen
  \bibfield  {author} {\bibinfo {author} {\bibfnamefont {P.~E.}\ \bibnamefont
  {Gaskell}}, \bibinfo {author} {\bibfnamefont {J.~J.}\ \bibnamefont {Thorn}},
  \bibinfo {author} {\bibfnamefont {S.}~\bibnamefont {Alba}}, \ and\ \bibinfo
  {author} {\bibfnamefont {D.~A.}\ \bibnamefont {Steck}},\ }\bibfield  {title}
  {\enquote {\bibinfo {title} {An open-source, extensible system for laboratory
  timing and control},}\ }\href {\doibase 10.1063/1.3250825} {\bibfield
  {journal} {\bibinfo  {journal} {Rev. Sci. Instrum.}\ }\textbf {\bibinfo
  {volume} {80}},\ \bibinfo {pages} {115103} (\bibinfo {year}
  {2009})}\BibitemShut {NoStop}%
\bibitem [{\citenamefont {Starkey}\ \emph {et~al.}(2013)\citenamefont
  {Starkey}, \citenamefont {Billington}, \citenamefont {Johnstone},
  \citenamefont {Jasperse}, \citenamefont {Helmerson}, \citenamefont {Turner},\
  and\ \citenamefont {Anderson}}]{Starkey2013}%
  \BibitemOpen
  \bibfield  {author} {\bibinfo {author} {\bibfnamefont {P.~T.}\ \bibnamefont
  {Starkey}}, \bibinfo {author} {\bibfnamefont {C.~J.}\ \bibnamefont
  {Billington}}, \bibinfo {author} {\bibfnamefont {S.~P.}\ \bibnamefont
  {Johnstone}}, \bibinfo {author} {\bibfnamefont {M.}~\bibnamefont {Jasperse}},
  \bibinfo {author} {\bibfnamefont {K.}~\bibnamefont {Helmerson}}, \bibinfo
  {author} {\bibfnamefont {L.~D.}\ \bibnamefont {Turner}}, \ and\ \bibinfo
  {author} {\bibfnamefont {R.~P.}\ \bibnamefont {Anderson}},\ }\bibfield
  {title} {\enquote {\bibinfo {title} {A scripted control system for autonomous
  hardware-timed experiments},}\ }\href {\doibase 10.1063/1.4817213} {\bibfield
   {journal} {\bibinfo  {journal} {Rev. Sci. Instrum.}\ }\textbf {\bibinfo
  {volume} {84}},\ \bibinfo {pages} {085111} (\bibinfo {year}
  {2013})}\BibitemShut {NoStop}%
\bibitem [{\citenamefont {Keshet}\ and\ \citenamefont
  {Ketterle}(2013)}]{Keshet2013}%
  \BibitemOpen
  \bibfield  {author} {\bibinfo {author} {\bibfnamefont {A.}~\bibnamefont
  {Keshet}}\ and\ \bibinfo {author} {\bibfnamefont {W.}~\bibnamefont
  {Ketterle}},\ }\bibfield  {title} {\enquote {\bibinfo {title} {A distributed,
  graphical user interface based, computer control system for atomic physics
  experiments},}\ }\href {\doibase 10.1063/1.4773536} {\bibfield  {journal}
  {\bibinfo  {journal} {Rev. Sci. Instrum.}\ }\textbf {\bibinfo {volume}
  {84}},\ \bibinfo {pages} {015105} (\bibinfo {year} {2013})}\BibitemShut
  {NoStop}%
\bibitem [{\citenamefont {Pruttivarasin}\ and\ \citenamefont
  {Katori}(2015)}]{Pruttivarasin2015}%
  \BibitemOpen
  \bibfield  {author} {\bibinfo {author} {\bibfnamefont {T.}~\bibnamefont
  {Pruttivarasin}}\ and\ \bibinfo {author} {\bibfnamefont {H.}~\bibnamefont
  {Katori}},\ }\bibfield  {title} {\enquote {\bibinfo {title} {Compact field
  programmable gate array-based pulse-sequencer and radio-frequency generator
  for experiments with trapped atoms},}\ }\href {\doibase 10.1063/1.4935476}
  {\bibfield  {journal} {\bibinfo  {journal} {Rev. Sci. Instrum.}\ }\textbf
  {\bibinfo {volume} {86}},\ \bibinfo {pages} {115106} (\bibinfo {year}
  {2015})}\BibitemShut {NoStop}%
\bibitem [{\citenamefont {Bourdeauducq}\ \emph {et~al.}(2018)\citenamefont
  {Bourdeauducq} \emph {et~al.}}]{artiq2018}%
  \BibitemOpen
  \bibfield  {author} {\bibinfo {author} {\bibfnamefont {S.}~\bibnamefont
  {Bourdeauducq}} \emph {et~al.},\ }\href {\doibase 10.5281/zenodo.1492176}
  {\enquote {\bibinfo {title} {m-labs/artiq: 4.0},}\ } (\bibinfo {year}
  {2018})\BibitemShut {NoStop}%
\bibitem [{\citenamefont {Perego}\ \emph {et~al.}(2018)\citenamefont {Perego},
  \citenamefont {Pomponio}, \citenamefont {Detti}, \citenamefont {Duca},
  \citenamefont {Sias},\ and\ \citenamefont {Calosso}}]{Perego2018}%
  \BibitemOpen
  \bibfield  {author} {\bibinfo {author} {\bibfnamefont {E.}~\bibnamefont
  {Perego}}, \bibinfo {author} {\bibfnamefont {M.}~\bibnamefont {Pomponio}},
  \bibinfo {author} {\bibfnamefont {A.}~\bibnamefont {Detti}}, \bibinfo
  {author} {\bibfnamefont {L.}~\bibnamefont {Duca}}, \bibinfo {author}
  {\bibfnamefont {C.}~\bibnamefont {Sias}}, \ and\ \bibinfo {author}
  {\bibfnamefont {C.~E.}\ \bibnamefont {Calosso}},\ }\bibfield  {title}
  {\enquote {\bibinfo {title} {A scalable hardware and software control
  apparatus for experiments with hybrid quantum systems},}\ }\href {\doibase
  10.1063/1.5049120} {\bibfield  {journal} {\bibinfo  {journal} {Rev. Sci.
  Instrum.}\ }\textbf {\bibinfo {volume} {89}},\ \bibinfo {pages} {113116}
  (\bibinfo {year} {2018})}\BibitemShut {NoStop}%
\bibitem [{\citenamefont {Donnellan}\ \emph {et~al.}(2019)\citenamefont
  {Donnellan}, \citenamefont {Hill}, \citenamefont {Bowden},\ and\
  \citenamefont {Hobson}}]{Donnellan2019}%
  \BibitemOpen
  \bibfield  {author} {\bibinfo {author} {\bibfnamefont {S.}~\bibnamefont
  {Donnellan}}, \bibinfo {author} {\bibfnamefont {I.~R.}\ \bibnamefont {Hill}},
  \bibinfo {author} {\bibfnamefont {W.}~\bibnamefont {Bowden}}, \ and\ \bibinfo
  {author} {\bibfnamefont {R.}~\bibnamefont {Hobson}},\ }\bibfield  {title}
  {\enquote {\bibinfo {title} {A scalable arbitrary waveform generator for
  atomic physics experiments based on field- programmable gate array
  technology},}\ }\href {\doibase 10.1063/1.5051124} {\bibfield  {journal}
  {\bibinfo  {journal} {Rev. Sci. Instrum.}\ }\textbf {\bibinfo {volume}
  {90}},\ \bibinfo {pages} {043101} (\bibinfo {year} {2019})}\BibitemShut
  {NoStop}%
\bibitem [{Sch()}]{SchreckCS_URL}%
  \BibitemOpen
  \href@noop {} {}\bibinfo {note} {Group of F. Schreck (Amsterdam)
  \url{http://www.strontiumbec.com/Control/Control.html}.}\BibitemShut {Stop}%
\bibitem [{LEN()}]{LENS_URL}%
  \BibitemOpen
  \href@noop {} {}\bibinfo {note} {Electronic shop at LENS (Florence)
  \url{http://ew.lens.unifi.it/}.}\BibitemShut {Stop}%
\bibitem [{\citenamefont {Canuel}\ \emph {et~al.}(2018)\citenamefont {Canuel},
  \citenamefont {Bertoldi}, \citenamefont {Amand}, \citenamefont {di~Borgo},
  \citenamefont {Chantrait}, \citenamefont {Danquigny}, \citenamefont
  {{\'{A}}lvarez}, \citenamefont {Fang}, \citenamefont {Freise}, \citenamefont
  {Geiger}, \citenamefont {Gillot}, \citenamefont {Henry}, \citenamefont
  {Hinderer}, \citenamefont {Holleville}, \citenamefont {Junca}, \citenamefont
  {Lef{\`{e}}vre}, \citenamefont {Merzougui}, \citenamefont {Mielec},
  \citenamefont {Monfret}, \citenamefont {Pelisson}, \citenamefont
  {Prevedelli}, \citenamefont {Reynaud}, \citenamefont {Riou}, \citenamefont
  {Rogister}, \citenamefont {Rosat}, \citenamefont {Cormier}, \citenamefont
  {Landragin}, \citenamefont {Chaibi}, \citenamefont {Gaffet},\ and\
  \citenamefont {Bouyer}}]{Canuel2018}%
  \BibitemOpen
  \bibfield  {author} {\bibinfo {author} {\bibfnamefont {B.}~\bibnamefont
  {Canuel}}, \bibinfo {author} {\bibfnamefont {A.}~\bibnamefont {Bertoldi}},
  \bibinfo {author} {\bibfnamefont {L.}~\bibnamefont {Amand}}, \bibinfo
  {author} {\bibfnamefont {E.~P.}\ \bibnamefont {di~Borgo}}, \bibinfo {author}
  {\bibfnamefont {T.}~\bibnamefont {Chantrait}}, \bibinfo {author}
  {\bibfnamefont {C.}~\bibnamefont {Danquigny}}, \bibinfo {author}
  {\bibfnamefont {M.~D.}\ \bibnamefont {{\'{A}}lvarez}}, \bibinfo {author}
  {\bibfnamefont {B.}~\bibnamefont {Fang}}, \bibinfo {author} {\bibfnamefont
  {A.}~\bibnamefont {Freise}}, \bibinfo {author} {\bibfnamefont
  {R.}~\bibnamefont {Geiger}}, \bibinfo {author} {\bibfnamefont
  {J.}~\bibnamefont {Gillot}}, \bibinfo {author} {\bibfnamefont
  {S.}~\bibnamefont {Henry}}, \bibinfo {author} {\bibfnamefont
  {J.}~\bibnamefont {Hinderer}}, \bibinfo {author} {\bibfnamefont
  {D.}~\bibnamefont {Holleville}}, \bibinfo {author} {\bibfnamefont
  {J.}~\bibnamefont {Junca}}, \bibinfo {author} {\bibfnamefont
  {G.}~\bibnamefont {Lef{\`{e}}vre}}, \bibinfo {author} {\bibfnamefont
  {M.}~\bibnamefont {Merzougui}}, \bibinfo {author} {\bibfnamefont
  {N.}~\bibnamefont {Mielec}}, \bibinfo {author} {\bibfnamefont
  {T.}~\bibnamefont {Monfret}}, \bibinfo {author} {\bibfnamefont
  {S.}~\bibnamefont {Pelisson}}, \bibinfo {author} {\bibfnamefont
  {M.}~\bibnamefont {Prevedelli}}, \bibinfo {author} {\bibfnamefont
  {S.}~\bibnamefont {Reynaud}}, \bibinfo {author} {\bibfnamefont
  {I.}~\bibnamefont {Riou}}, \bibinfo {author} {\bibfnamefont {Y.}~\bibnamefont
  {Rogister}}, \bibinfo {author} {\bibfnamefont {S.}~\bibnamefont {Rosat}},
  \bibinfo {author} {\bibfnamefont {E.}~\bibnamefont {Cormier}}, \bibinfo
  {author} {\bibfnamefont {A.}~\bibnamefont {Landragin}}, \bibinfo {author}
  {\bibfnamefont {W.}~\bibnamefont {Chaibi}}, \bibinfo {author} {\bibfnamefont
  {S.}~\bibnamefont {Gaffet}}, \ and\ \bibinfo {author} {\bibfnamefont
  {P.}~\bibnamefont {Bouyer}},\ }\bibfield  {title} {\enquote {\bibinfo {title}
  {Exploring gravity with the {MIGA} large scale atom interferometer},}\ }\href
  {\doibase 10.1038/s41598-018-32165-z} {\bibfield  {journal} {\bibinfo
  {journal} {Sci. Rep.}\ }\textbf {\bibinfo {volume} {8}},\ \bibinfo {pages}
  {14064} (\bibinfo {year} {2018})}\BibitemShut {NoStop}%
\bibitem [{\citenamefont {Naik}\ \emph {et~al.}(2018)\citenamefont {Naik},
  \citenamefont {Kuyumjyan}, \citenamefont {Pandey}, \citenamefont {Bouyer},\
  and\ \citenamefont {Bertoldi}}]{Naik2018}%
  \BibitemOpen
  \bibfield  {author} {\bibinfo {author} {\bibfnamefont {D.~S.}\ \bibnamefont
  {Naik}}, \bibinfo {author} {\bibfnamefont {G.}~\bibnamefont {Kuyumjyan}},
  \bibinfo {author} {\bibfnamefont {D.}~\bibnamefont {Pandey}}, \bibinfo
  {author} {\bibfnamefont {P.}~\bibnamefont {Bouyer}}, \ and\ \bibinfo {author}
  {\bibfnamefont {A.}~\bibnamefont {Bertoldi}},\ }\bibfield  {title} {\enquote
  {\bibinfo {title} {Bose{\textendash}{E}instein condensate array in a
  malleable optical trap formed in a traveling wave cavity},}\ }\href {\doibase
  10.1088/2058-9565/aad48e} {\bibfield  {journal} {\bibinfo  {journal} {Quantum
  Sci. Technol.}\ }\textbf {\bibinfo {volume} {3}},\ \bibinfo {pages} {045009}
  (\bibinfo {year} {2018})}\BibitemShut {NoStop}%
\bibitem [{Meg()}]{Mega}%
  \BibitemOpen
  \href@noop {} {}\bibinfo {note} {To avoid confusion, we will always use
  {$\mathrm{M}=2^{20}$} and never {$\mathrm{M}=10^6$} when referring to memory
  size and number of instructions.}\BibitemShut {Stop}%
\bibitem [{Int()}]{IntMod}%
  \BibitemOpen
  \href@noop {} {}\bibinfo {note} {We make a distinction between a master
  module and an intelligent one since in a distributed architecture many
  auxiliary modules are intelligent.}\BibitemShut {Stop}%
\bibitem [{\citenamefont {Prevedelli}\ \emph {et~al.}(2019)\citenamefont
  {Prevedelli} \emph {et~al.}}]{zenodo2019}%
  \BibitemOpen
  \bibfield  {author} {\bibinfo {author} {\bibfnamefont {M.}~\bibnamefont
  {Prevedelli}} \emph {et~al.},\ }\href {\doibase 10.5281/zenodo.3571496}
  {\enquote {\bibinfo {title} {Yet another control system for {AMO} physics},}\
  } (\bibinfo {year} {2019})\BibitemShut {NoStop}%
\bibitem [{\citenamefont {Schreck}(2002)}]{SchreckPhD}%
  \BibitemOpen
  \bibfield  {author} {\bibinfo {author} {\bibfnamefont {F.}~\bibnamefont
  {Schreck}},\ }\emph {\bibinfo {title} {Mixtures of ultracold gases: {F}ermi
  sea and {B}ose-{E}instein condensate of {L}ithium isotopes}},\ \href@noop {}
  {Ph.D. thesis},\ \bibinfo  {school} {Univ. Pierre et Marie Curie - Paris VI}
  (\bibinfo {year} {2002})\BibitemShut {NoStop}%
\bibitem [{zte()}]{ztexURL}%
  \BibitemOpen
  \href@noop {} {}\bibinfo {note} {{ZTEX FPGA} {B}oard with {O}pen {S}ource
  {SDK}, {M}odels 2.04b and 2.13a
  \href{http://www.ztex.de}{www.ztex.de}.}\BibitemShut {Stop}%
\bibitem [{red()}]{redPitayaURL}%
  \BibitemOpen
  \href@noop {} {}\bibinfo {note} {{STEM}lab 125-14, formerly {R}ed {P}itaya,
  \href{http://redpitaya.com/}{redpitaya.com}.}\BibitemShut {Stop}%
\bibitem [{SDC()}]{SDCC}%
  \BibitemOpen
  \href@noop {} {}\bibinfo {note} {{Avaliable at}
  \href{https://sdcc.sourceforge.net}{sdcc.sourceforge.net}.}\BibitemShut
  {Stop}%
\bibitem [{ISE()}]{ISE}%
  \BibitemOpen
  \href@noop {} {}\bibinfo {note} {{Avaliable at}
  \href{https://www.xilinx.com/products/design-tools/ise-design-suite/ise-webpack.html}{www.xilinx.com/products/design-tools/ise-design-suite/ise-webpack.html}.}\BibitemShut
  {Stop}%
\bibitem [{ope()}]{opencores}%
  \BibitemOpen
  \href@noop {} {}\bibinfo {note} {{See}
  \href{https://opencores.org}{opencores.org}.}\BibitemShut {Stop}%
\bibitem [{gtk()}]{gtkwave}%
  \BibitemOpen
  \href@noop {} {}\bibinfo {note} {See
  \href{https://gtkwave.sourceforge.net/}{gtkwave.sourceforge.net}.}\BibitemShut
  {Stop}%
\bibitem [{vcd(2001)}]{vcd}%
  \BibitemOpen
  \bibfield  {title} {\enquote {\bibinfo {title} {{IEEE Standard Verilog
  Hardware Description Language}},}\ }\href {\doibase
  10.1109/IEEESTD.2001.93352} {\bibfield  {journal} {\bibinfo  {journal} {{IEEE
  Std 1364-2001}}\ ,\ \bibinfo {pages} {1--792}} (\bibinfo {year}
  {2001})}\BibitemShut {NoStop}%
\bibitem [{AD9()}]{AD9958_ds}%
  \BibitemOpen
  \href@noop {} {}\bibinfo {note} {{Available at}
  \href{https://www.analog.com/media/en/technical-documentation/data-sheets/AD9958.pdf}{www.analog.com/media/en/technical-documentation/data-sheets/AD9958.pdf}.}\BibitemShut
  {Stop}%
\bibitem [{FX3()}]{FX3}%
  \BibitemOpen
  \href@noop {} {}\bibinfo {note} {{Avaliable at}
  \href{https://www.cypress.com/documentation/software-and-drivers/ez-usb-fx3-software-development-kit}{www.cypress.com/documentation/software-and-drivers/ez-usb-fx3-software-development-kit}.}\BibitemShut
  {Stop}%
\bibitem [{Viv()}]{Vivado}%
  \BibitemOpen
  \href@noop {} {}\bibinfo {note} {{Available at}
  \href{https://www.xilinx.com/products/design-tools/vivado/vivado-webpack.html}{www.xilinx.com/products/design-tools/vivado/vivado-webpack.html}.}\BibitemShut
  {Stop}%
\end{thebibliography}%

\end{document}